# Lithium storage in titania films as a function of position: Unification of intercalation electrode and super-capacitor concepts

Chuanlian Xiao[1], Hongguang Wang[1], Peter A. van Aken[1], Robert Usiskin[1], Joachim Maier[1]*

[1]Max Planck Institute for Solid State Research, Stuttgart, Germany.

*office-maier@fkf.mpg.de

**Abstract**

We carefully investigated the storage of lithium in titania films on various substrates as a function of thickness. The experiments enable us to precisely separate contributions from bulk and boundary storage. The battery capacity measurements are complemented by bias dependent measurements of impedance, yielding interfacial resistance as well as interfacial capacitance. Independent information on electron and Li distribution is gained by scanning transmission electron microscopy (STEM), electron energy loss spectroscopy (EELS), aberration-corrected annular-bright-field (ABF) STEM. As a result, we obtain the full picture in terms of equilibrium storage (lithium content) and charge carrier concentrations as a function of spatial coordinates with cell voltage as a parameter. More importantly, both bulk storage which obeys electroneutrality and boundary storage which follows the space charge picture can be traced back to a common thermodynamic conception, and are obtained from it as special cases. This corresponds to no less than a unification of intercalation storage and super-capacitive storage, which are usually considered as independent phenomena, the reason for this lying in the hitherto lack of an adequate defect-chemical and nanoionic picture.

Lithium intercalation batteries belong certainly to the most important device developments of the last decades. They are indispensable for nowadays' energy household. They are particularly useful, whenever electro-mobility is concerned, be it powering mobile devices such as lap-taps or be it enabling mobility such as for battery-driven automobiles (*1-3*). Owing to the large capacities, they are also candidates for grid-storage. Not least for this enormous practical relevance, the 2019 Chemistry Nobel Prize was conferred for the invention of the rocking chair battery based on Li-cobaltate and graphite as intercalation electrodes (*4, 5*). In spite of all of this, the charge carrier situation (point defect chemistry) of electrodes is not properly addressed (*6*).

Intercalation allows the entire volume of the storage electrode to be chemically capacitive (see Fig.1). The electrode is both ionically and electronically conductive. The consequential huge capacity gain when compared to interfacial effects is simultaneously the reason why this goes along with a comparatively low power density as the necessary chemical diffusion process slows down the (dis-)charging rate.

The opposite is true for supercapacitors (7). Here, the charging is restricted to the interface and no bulk chemical diffusion is necessary, leading to higher power densities but lower energies (8-11) (see Fig. 1). This field of application has been explored with a similar intensity though typically by a different community. As a consequence, the two storage modes appear to be more separated than they really deserve. We show that this is a consequence of insufficient understanding of the charge carrier chemistry in the material under concern.

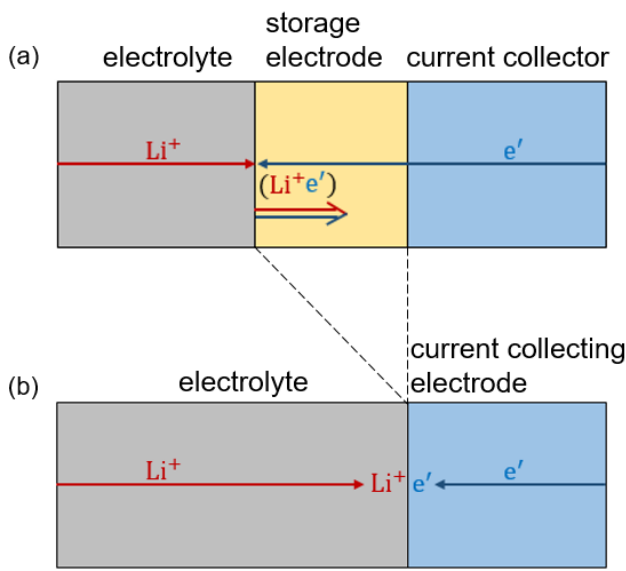

Fig. 1 Sketch of (a) bulk storage (in a predominantly electronically conducting electrode) and (b) supercapacitive storage (at electrolyte/current collecting electrode interface). In the general case of a mixed conductor, both processes occur simultaneously.

Supercapacitors in the original sense fully rely on electrical polarization. Two antagonistic double layers are formed at the two electrode-electrolyte interfaces and overall there is mass conservation within the electrolyte. Here, the electrode acts purely as an electronic conductor. Even though not the topic of this contribution, it is to be mentioned that there are variants of supercapacitive storage where a Faradaic electrode reaction occurs, but still only the interface is affected (12-16). Such pseudo-capacitors may even have higher capacities, and the term electrochemical capacitor has been used as generic term (7). One can also employ a reversible electrode on one side such that mass changes occur at the expense of the reversible side and the lithium potential is varied on the other side on varying the reversible cell voltage.

A different, more conceptual and more general viewpoint is taken if the ensemble of the electronically conducting and the ionically conducting phases is considered as a composite electrode (6, 17). Then excess storage can be viewed therein in a similar way as it was discussed for composite electrolytes in terms of transport (18). A composite of an electron conductor with no possibility to accommodate a Li-ion and an ion conductor that cannot be hosting the electron (e.g. no transition metal), can now store lithium ($Li^+ + e^-$) in a job-sharing way. That such storage can be substantial and quick has been evidenced for various examples including $Li_2O$:Ru, LiF:Ni and RbAg₄I₅:C (*19-22*). While the vast majority of evidence was derived from impedance measurements and elucidation of the activity dependence, recently direct magnetometric evidence was reported (*23, 24*). In the case of RbAg₄I₅:C also a deficiency can be realized and a charge-discharge curve fully covering the deficient and the excess part of the storage can be measured. In fact, if the sizes involved are mesoscopic, one should be able to realize artificial electrodes (analogously to artificial electrolytes as obtained in the nanoionics CaF$_2$-BaF$_2$ system (*25*)) with the potential of even overcoming the dichotomy between energy and power density (*17*). If the ion conducting second phase can be identified with the electrolyte of an electrochemical cell, the set-up is the one of a super-capacitor cell with a reversible electrode (Fig. 1). Yet, the job-sharing conception is thermodynamically based and more general. This may be most obvious in the predicted finding that the Li$_2$O-Ru composite can also synergistically store hydrogen (*26*) via such job-sharing. More importantly, this conception puts us into the position to treat the storage thermodynamics and kinetics within the framework of point-defect chemistry.

Most significantly – and this is the essence of the present paper – the understanding of bulk and boundary charge carrier chemistry allows us to *unify the two conceptions (intercalation and supercapacitive function) by considering equilibrium storage as a function of position* (distance from the interface), whereby the bulk values reflect the intercalation electrode function and the boundary values the supercapacitive function.

We use TiO$_2$-films as the subject of our studies and consider the local storage capacity by measuring the lithium uptake as a function of film thickness for various electron-accepting substrates. Bulk TiO$_2$ in anatase or rutile form can host lithium (*27, 28*), but - as we show – in addition the interface to an electron conductor such as doped SrTiO$_3$ (ST) or Ru can take up an excess beyond the bulk value through a space charge effect. The possibility of interfacial storage in anatase is also obvious from studies of nanoparticles at different rates (*13, 29*).

Analogously to thin film studies of parallel conductances as functions of thickness, the study of capacitance as function of thickness allows us to separate bulk and boundary effects. In equilibrium, one expects a straight line with a slope that indicates the bulk capacity increment and an intercept that gives the excess boundary value. Adopting the usual colloquial notion of expressing the battery capacity by the stored charge (Q), the area (a) specific capacity depends linearly on the thickness (L) as

$Q/a = Q_{int}/a + q_{bulk}L = Q_{int}/a + c_{bulk}FL = Q_{int}/a + (\delta/V)FL$,

whereby $q_{bulk}$ denotes the neutral bulk charge density that is given by the L-independent bulk defect concentration $\delta/V$ where $\delta$ (=$n_{Li}/n_{Ti}$) denotes the molar fraction and V the molar volume.

The titania modification that we refer to in the thin films is anatase, the defect chemistry of which is well-understood (*27, 29, 30*). In particular, it has been shown that Li ions occupy interstitial positions in the bulk up to a concentration of at least $\delta$ =0.5 ($Li_\delta TiO_2$) compensated by conduction electrons in the Ti d orbitals (Ti$^{3+}$) (*31*).

Our analysis of electrochemical storage experiments is accompanied by conductance and capacitance across the interfaces under regard with and without bias as well as on electron microscopy. The electrical set-ups are sketched in Fig. 2. Not only do the results unambiguously show that in addition to bulk storage, significant excess interfacial storage effects occur, which can be ascribed to job-sharing and are restricted to the space charge zones. They also allow us to plot the local amount of lithium (i.e. stored charge) as a function of position with the OCV voltage (Li-activity) as parameter. The results will be quantitatively interpreted in the light of the defect-chemical and nanoionics concepts (6), and the full picture unifying the intercalation and supercapacitive viewpoints can be obtained.

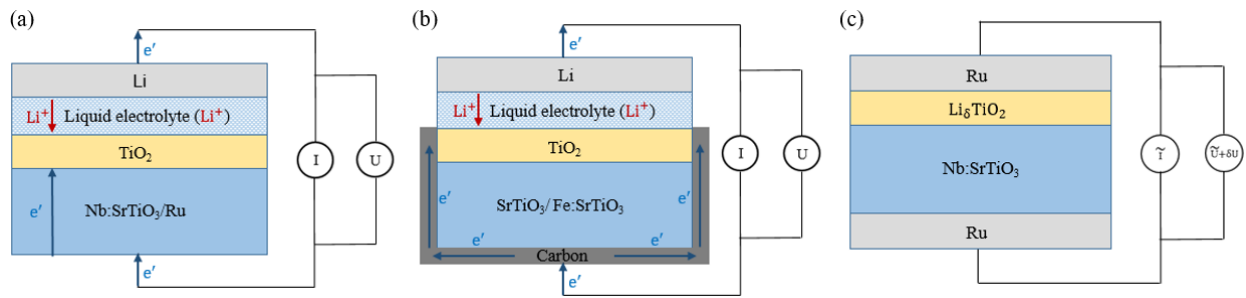

Fig. 2 Measurement set-ups: (a) storage experiment with a conductive substrate (b) storage experiment with a non-conductive substrate and (c) impedance measurement (perpendicular to the interface).

Let us consider the results step by step. Fig. 3 gives a compilation of representative storage experiments of titania thin films with various substrates as a function of thickness, which we precisely determined using various methods (x-ray reflectivity, TEM, stylus profilometry). The Li-uptake of a dummy cell has been carefully measured and subtracted. The capacity values were measured accurately under constant current. The extrapolation of the capacity values to zero thicknesses yields the interfacial contribution.

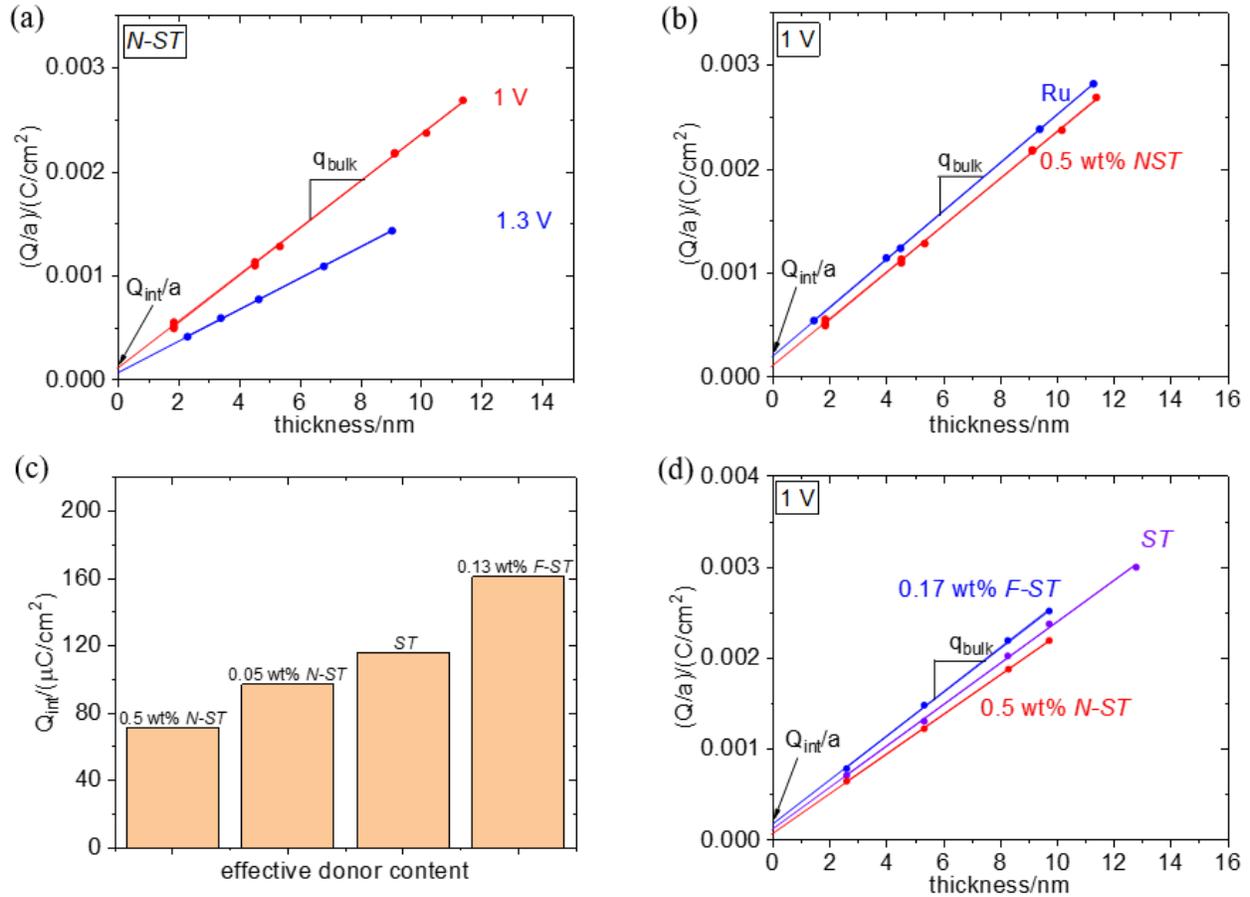

Fig. 3 Storage measurement: (a) Capacity ($Q/a$) as a function of thickness at different degrees of storage (TiO$_2$ on 0.5 wt% N-ST substrates). The bulk capacities (the slopes) are $2.26 \times 10^9$, $1.50 \times 10^9$ $\mu C/cm^3$ and the interfacial storage capacities (the intercepts) are 92 and 72 $\mu C/cm^2$ in the case of 1 V and 1.3 V, respectively. (b) Capacity ($Q/a$) as a function of thickness for different substrates (Ru and 0.5 wt% N-ST substrates) at 1 V. The bulk capacities (the slopes) are $2.31 \times 10^9$, $2.26 \times 10^9$ $\mu C/cm^3$ and the interfacial storage capacities (the intercepts) are 203 and 92 $\mu C/cm^2$ for Ru and 0.5 wt% N-ST, respectively. (c) Interfacial storage capacity ($Q_{int}/a$) as a function of effective donor content. (d) Capacity ($Q/a$) as a function of thickness for differently doped substrates (0.13 wt% F-ST, ST and 0.5 wt% N-ST substrates) at 1V. The bulk capacities (the slopes) are $2.40 \times 10^9$, $2.27 \times 10^9$, $2.20 \times 10^9$ $\mu C/cm^3$ and the interfacial storage capacities (the intercepts) are 161, 116 and 71 $\mu C/cm^2$ for 0.13% F-ST, ST and 0.5 wt% N-ST substrates, respectively. TiO$_2$ films were deposited by atomic layer deposition (ALD) (a), (b) and pulsed laser deposition (PLD), molecular beam epitaxy (MBE) (c), (d).

The measured bulk and boundary values correspond to typical values for titania (*13, 31, 32*). The boundary values which are on the order of $\sim 100\ \mu C/cm^2$, are - though rather high - not inconsistent with monolayer values. As a consistency check let us assume the limit of a fully occupied first layer in titania within a discrete picture (*33*). As two interstitial sites per Ti are available (*26*), we estimate a maximum of $850\ \mu C/cm^2$. In fact, the limiting factor may be to accommodate the equivalent amount of electrons in the substrate. On a similarly rough level of approximation we assume in the first layer of SrTiO₃, a maximum of one electron per Ti, we estimate a limit of 105 $\mu C/cm^2$. In fact, the screening length is a few to tens of nanometers for *ST* substrate. Then the excess stored charge could be $\sim 140\ \mu C/cm^2$ if a Gouy-Chapman profile is applied. Although none of the measured interfacial values exceeds the order of such a limit, partial association (of Li⁺ and e⁻) is probable (*13, 27*) ("pseudo-capacitive contributions") and would make the rather high values more expectable.

It is most important in our context, that we are not only able to precisely deconvolute bulk and boundary storage, but that they can be traced back to a common thermodynamic framework, namely the generalized defect-chemical concept (*18*), expressed by the spatial profile of the ionic and electronic charge carriers. The profile $c_{Li}$ can be quantitatively given as

$$c_{Li} = fct(x, c_0, c_{bulk})$$

where both boundary and bulk concentration can be derived from the respective mass action laws. Both $c_0$ and $c_{bulk}$ depends in a predictable way on the control parameters temperature, pressure, doping content and Li-activity (i.e. OCV). As far as $c_{bulk}$ is concerned, electroneutrality has to be involved, and as far as $c_0$ is concerned the respective mass action law naturally involves also substrate parameters (electron accommodation). The total amount of stored lithium ($Q$) can be obtained by spatial integration. Integration of the excess $c_{Li}(x) - c_{bulk}$ yields $Q_{int}$ and the difference between $Q$ and $Q_{int}$ the bulk storage $Q_{bulk}$ ($\propto c_{bulk} L$). Here it suffices to consider the influence of the most important parameters.

Let us first consider the impact of different degrees of storage (voltage) on the results. (Fig. 3a gives the results for bulk (slope) and boundary values (intercept) for TiO₂ on a donor (Nb) doped SrTiO₃ substrate (*N-ST*) for two different degrees of storage). As bulk and boundary values depend differently on the Li activity, the ratio of the two storage capacities varies with Li content. In order to show that this coupling can be obtained from the thermodynamic framework, let us refer to small storage, i.e. to the dilute bulk thermodynamics (*6, 34*), and negligible electric potential drop over the interface. We expect power laws for Q as a function of the lithium activity (power: *N*) that are different for bulk ($N \geq 1/2$) and for boundary ($N \leq 1/4$) and hence a ratio of boundary and bulk storage that depends on the degree of storage. For dilute conditions, this ratio then follows a power law with a power smaller than -1/4. The *N*-values measured and displayed in Fig. 3a are clearly beyond the dilute limit (0.035 for the bulk and 0.021 for the boundary) and crystallographic and electronic saturation effects leading to a flattening of the $Q(a_{Li})$ curve are expected. Owing to the dilute exponents, saturation will set in earlier for the boundaries and the tendency for the ratio is expected to be reversed.

Now let us consider the effect of the electronic situation in the substrate. The mass action laws for the bulk storage show that for similar mass action constants (similar chemistry) the Li-uptake should decrease with increasing electron concentration already present in the material $c_{n\infty}$ as for the amount of active electrons $c_n(x) = c_{n\infty} + (Q(x)/F)$ (entropic effect). This is particularly relevant for interfacial storage using differently doped *ST* substrates. Then the capacity should increase from more highly donor doped *ST* to lowly doped, from donor doped to undoped, from undoped to acceptor doped (Fe doped SrTiO₃ (*F-ST*)). Again, this is impressively corroborated by Fig. 3c,d displaying the same sequence for the interfacial capacities. (Note that unlike donor doped *ST*, the electron conductivity of the pure or acceptor doped *ST* is poor and a slightly different measuring set-up had to be used (Fig. 2b).)

Unsurprisingly the highest interfacial storage capacities are obtained, if *ST* is replaced by Ru owing to its expectedly more favorable heterogeneous mass action constant, (i.e. essentially due to the lower energy for accepting excess electrons, owing to the noble character of the metal). (Regarding Ru substrate (hexagonal close-packed), the atom density is $1.58 \times 10^{15}$ cm$^{-2}$. The stored charge is 250 $\mu C/cm^2$ if each Ru atom at first layer takes up one electron.) The storage mass action constant is even more favorable, if one takes account of a monolayer of ruthenium oxide (for which there is evidence from XPS) ( such a monolayer will have a maximum capacity of 160 $\mu C/cm^2$ if each Ru takes up one electron).

Consequently, in the cell arrangement given in Fig. 2c for the perpendicular electric measurements the most resistive interface is the Ru/TiO₂ interface. The bias dependence derived from the bias dependence of the impedance, follows accurately the expected square root dependence before saturation effects become perceptible (Fig. 4c). As far as the evaluation of the interfacial electric capacitance (Fig. 4d) is concerned, various interfaces contribute and a rather low bias dependence is observed.

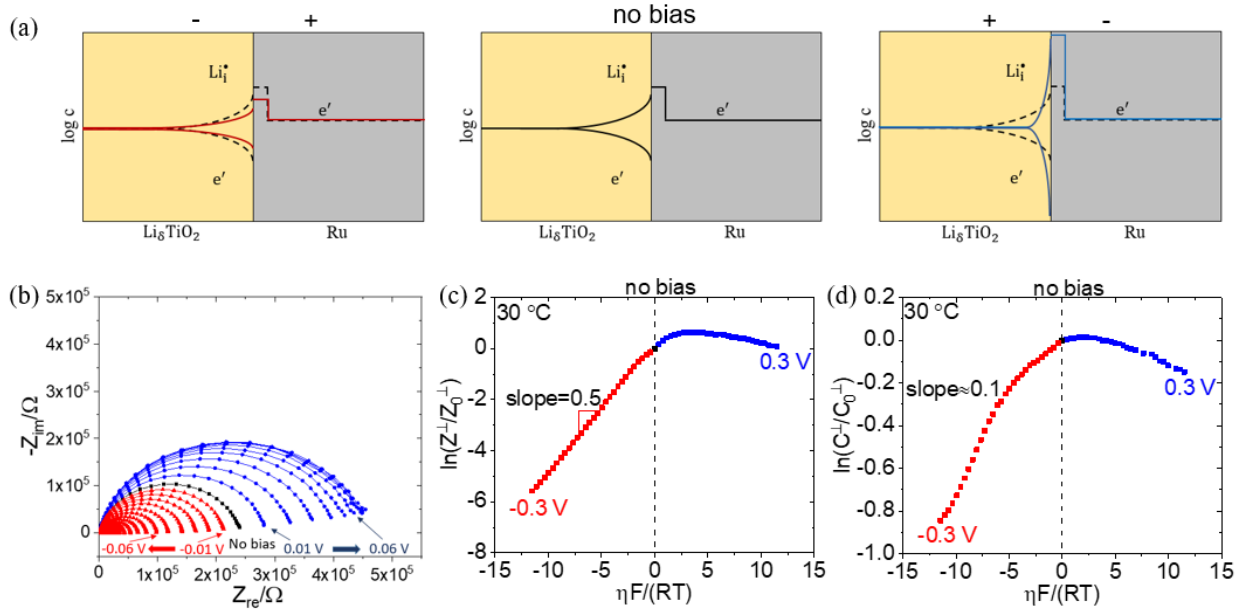

Fig. 4 Impedance measurement (perpendicular to the interface). (a) Charge carrier concentration profiles (cross the $Li_8TiO_2$/Ru interface) as a function of bias. (b) Impedance spectrum, (c) resistance and (d) capacitance as a function of bias.

Parallel measurements that filter out electronic and ionic contributions can be helpful as well but are experimentally very challenging.

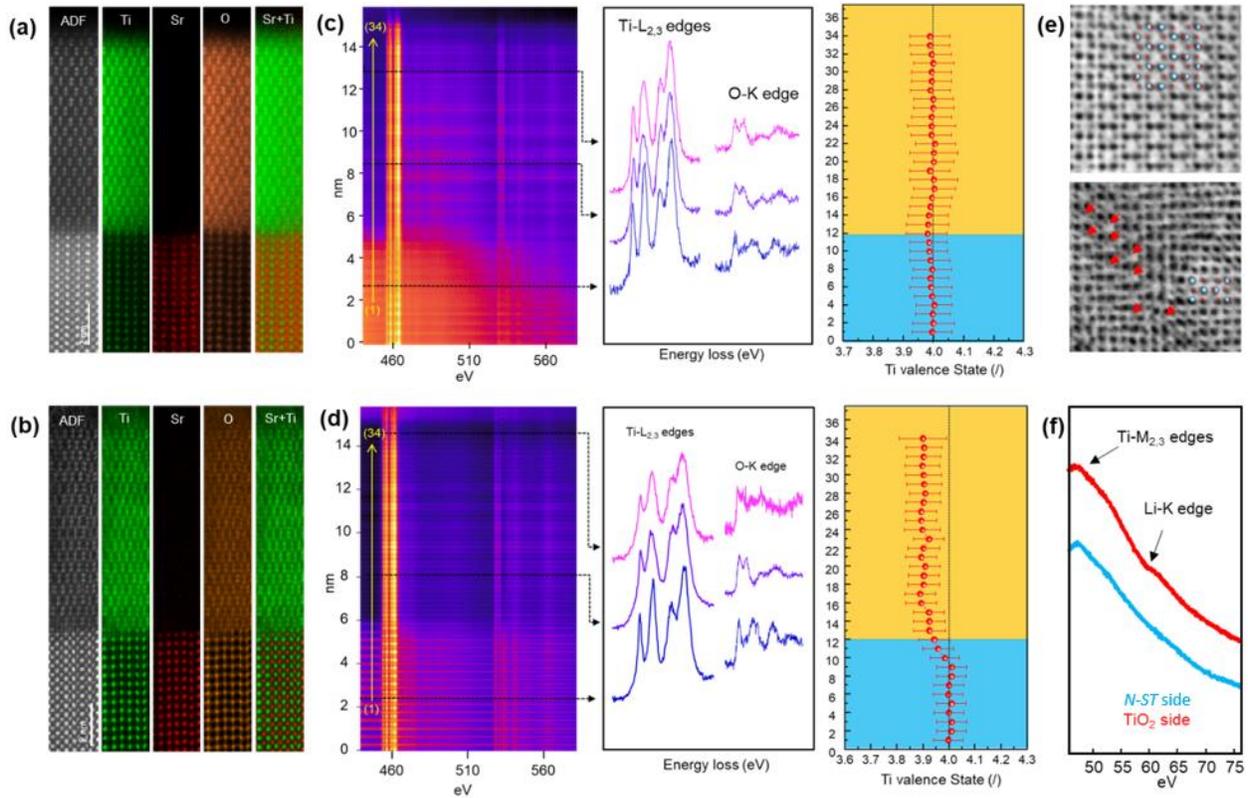

Fig. 5 STEM and EELS measurements. (a), (b) STEM-EELS elemental maps across the interface of $TiO_2$/*N-ST* and $Li_{0.1}TiO_2$/*N-ST*, respectively. (c), (d) EELS line scan across the interface of $TiO_2$/*N-ST* and $Li_{0.1}TiO_2$/*N-ST*, corresponding EELS spectra of Ti-$L_{2,3}$ and O-K edges, and fine structure analysis of Ti-$L_{2,3}$ edges. (e) ABF images of $TiO_2$ (upper panel) and $Li_{0.1}TiO_2$ (bottom panel). (f) EELS spectra of Li-K edge at the *N-ST* side and lithiated $TiO_2$ side. The doping content (Nb) of *ST* substrate is 0.5 wt%.

More information, in fact very independent and very clear corroboration, is obtained by scanning transmission electron microscopy in combination with electron energy-loss spectroscopy. Fig. 5a and Fig. 5b show that the grown $TiO_2$ film has good crystalline quality, which maintains after lithiation. Furthermore, the interfaces between $Li_\delta TiO_2$ and *N-ST* under consideration keep atomically sharp according to atomic-resolution maps of the constituent elements. The results of fine structure analysis of Ti-$L_{2,3}$ edges (Fig. 5c and 5d) exactly reflect the storage model: at low storage, the Ti valence in $TiO_2$ is +3.9 and increases near the boundary owing to the electronic depletion. On the *N-ST* side of the interface, the valence is lower than +4 close to the interface (electronic accumulation) and increases towards +4 when approaching the bulk. On the $TiO_2$ side, the interfacial effect is very short-ranged while it is on the order of 1 nm on the *N-ST* side. (In fact, the Debye screening length in 0.5 wt% *N-ST* is about 1.1 nm, which is consistent with the EELS results and the one in $Li_\delta TiO_2$ less than the atomic spacing.) Fig. 6 sketches the general case and compares it with pure bulk and pure interfacial storage. Note that for the sake of illustration, different scales are used for $Li_\delta TiO_2$ and *N-ST*.

In case of lithiation, the intercalated Li can be observed in the ABF image of $Li_\delta TiO_2$ (as indicated by red arrow in Fig. 5(e)) comparing to that of TiO₂. Furthermore, EELS spectra of Li-K edge unambiguously evidence the presence of Li signal in TiO₂ rather than in *N-ST* (Fig. 5f), consistent with ABF results.

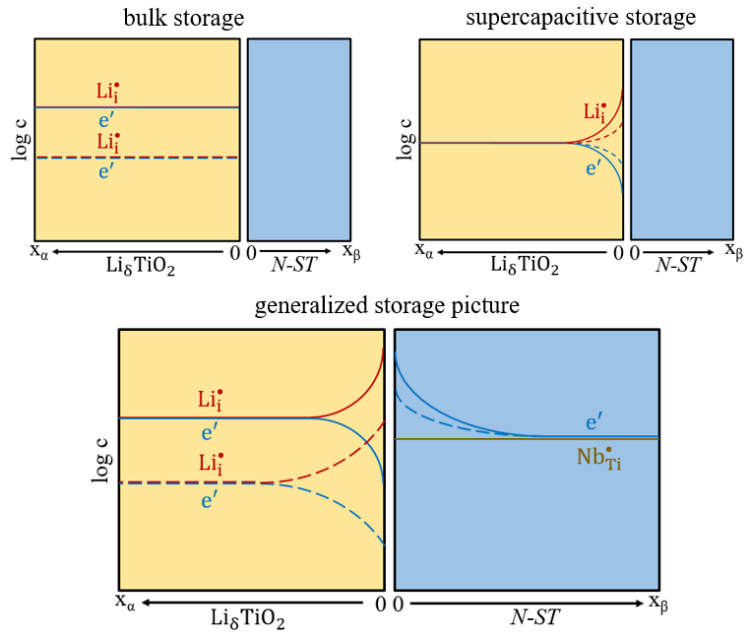

Fig. 6 The charge carrier concentration profiles for bulk storage, supercapacitive storage and generalized storage for TiO₂, here contacted with *N-ST* highlight the basic message of the contribution. For the sake of illustration, different scales are used for $x_\alpha$ and $x_\beta$. Owing to the much smaller Debye length in $Li_\delta TiO_2$, the space charge zone (not more than one unit cell even for the smaller $\delta \approx 0.1$) is much narrower than for *N-ST*.

In conclusion, we have analyzed the storage of lithium in a typical mixed conductor (TiO₂) as a function of position with the nature of the contact phase and the degree of storage as parameters. The results can be excellently understood in terms of the generalized charge carrier thermodynamics that takes account of the positional dependence of storage and hence includes bulk and boundary effects. Not only could we precisely deconvolute bulk and boundary contributions, we could also investigate their spatial dependences. Most importantly, we were able to derive the entire profile and thus bulk and boundary contributions from a common generalized thermodynamic model. Hence, this treatment unifies bulk intercalation and supercapacitive mechanisms. The first follows as a special case if interfacial storage is ignored (as it is a good approximation for macroscopic mixed conductors) and the second if electrons are not mobile in the bulk (as it is a good approximation for electrolyte materials). This generalization clearly demonstrates that on one hand the separation into two apparently very different phenomena (often treated in disjunct scientific communities) is neither adequate nor necessary, and that on the other hand the treatment indicates a way of reconciling the energy and power density conflict by varying size and contact phases.

**Acknowledgements**

We thank Georg Cristiani, Gennady Logvenov, Yuanshan Zhang, Julia Deuschle, Xu Chen, Rotraut Merkle, Davide Moia, Chia-Chin Chen, Kathrin Küster, Helga Hoier, Florian Kaiser, Thomas Reindl, Marion Hagel, Chuanhai Gan, Yue Zhu, Christian Berger, Peter Kopold, Armin Sorg, Annette Fuchs, Udo Klock, Uwe Traub, Barbara Baum for scientific discussions and technical assistance.